# Metalorganic Chemical Vapor Deposition of AlScN Thin Films and AlScN/AlN/GaN Heterostructures

Vijay Gopal Thirupakuzi Vangipuram[1], Abdul Mukit[1], Kaitian Zhang[1], Salva Salmani-Rezaie[2], Hongping Zhao[1,2,*]

[1] Department of Electrical and Computer Engineering, The Ohio State University, Columbus, Ohio 43210

[2] Department of Materials Science and Engineering, The Ohio State University, Columbus, Ohio 43210

[*]Corresponding author: zhao.2592@osu.edu

## Abstract

AlScN thin films were grown via metalorganic chemical vapor deposition (MOCVD), showing controllable incorporation of scandium (Sc) into the AlN lattices. Systematic variation of growth parameters demonstrated an obvious influence on Sc incorporation, with X-ray photoelectron spectroscopy (XPS) analysis indicating Sc composition up to ~13% when $(MCp)_2ScCl$ was used as the precursor. AlScN/AlN/GaN heterostructures grown on GaN templates exhibited the formation of a two-dimensional electron gas (2DEG) channel at the AlScN/AlN–GaN interface, confirming their potential use in high electron mobility transistor (HEMT) device technologies. Variation in AlScN/AlN barrier thickness within the heterostructures showed that thicker barriers yield higher sheet charge densities from both Hall and capacitance-voltage (C-V) measurements. With an AlScN/AlN barrier thickness of ~30 nm, sheet charge density of $5.22 \times 10^{12}$ cm$^{-2}$ was extracted from C-V. High-resolution scanning transmission electron microscopy (S/TEM) further confirmed Sc incorporation and revealed the wurtzite crystalline structure of the films and heterostructures. These results establish MOCVD growth of AlScN as a promising and compatible material for advancing III-nitride heterostructures in high performance electronics and potentially ferroelectrics.

## Introduction

Over the past few decades, the III-nitride material system, consisting of AlN, InN, and GaN, has enabled significant advancements and widespread adoption in high-frequency, high-power, and optoelectronic applications. The ability to form alloys among these binary compounds, all sharing the same wurtzite crystal structure, has provided extensive flexibility for designing high-performance electronic and optoelectronic devices. The realization of high-quality III-nitride epitaxial layers through metalorganic chemical vapor deposition (MOCVD) has been pivotal in advancing and commercializing modern device technologies, including light emitting diodes (LEDs), laser diodes, and high electron mobility transistors (HEMTs).[1,2]

AlScN has several advantageous material properties that show great promise in further developing novel devices and expanding the available application spaces where the wurtzite nitride materials can be utilized. Incorporation of Sc has shown significant increase in the piezoelectric response as compared to AlN; providing opportunities within micro-electromechanical systems (MEMS) resonators, radio frequency (RF) filters and acoustic devices.[3–7] Moreover, the discovery of ferroelectricity within AlScN proves to be a key finding in garnering significant interest in further developing the material towards memory applications.[8,9] In addition, incorporating Sc has proven to also show great promise in bandgap engineering for optical and optoelectronic devices.[10,11] A key advantage of AlScN, alloyed at Sc composition between ~9-14% can form a lattice-matched structure to GaN.[9,12–14] This allows for epitaxy of thicker layers while still providing a significant band-offset with GaN. The opportunity to independently vary barrier thickness while maintaining a large band-offset with GaN provides possibility for designing next-generation HEMT device structures. This contrasts with the traditional AlGaN/GaN heterostructures, where the strain of the heterostructure and the band-offset achieved through compositional tuning have a known trade-off.

While AlScN thin film deposition and epitaxial growth have been widely studied by sputtering and molecular beam epitaxy (MBE); there has been one primary group who has demonstrated high quality AlScN thin films via MOCVD. In their work, researchers at Fraunhofer Institute IAF used a customized close-coupled showerhead MOCVD reactor to achieve Sc incorporation at compositional levels up to 30%.[15] A key limiting factor in achieving Sc incorporation via MOCVD is the lack of a suitable metalorganic (MO) precursor that can provide sufficient vapor pressure required for MOCVD growth. To circumvent this limitation, the MO precursor and delivery lines require high-temperature heating.[15–17]

**Experimental Details**

In this study, all films described, unless otherwise stated, were grown on c-plane GaN-on-sapphire templates using a SiC-coated graphite susceptor. The MOCVD system was modified to enable high-temperature delivery of the Sc-containing MO precursor to the reactor. As previously stated, it is crucial that the Sc source is heated to achieve sufficient vapor pressures that can be comparable with vapor pressures achieved for the Al-based precursor. This is required to ensure composition levels of Sc incorporation within the grown films. In this work, trimethylaluminum (TMAl) was used as the Al precursor, while $NH_3$ was used as the precursor for the N-species. Hydrogen ($H_2$) was utilized as a carrier gas for all conditions explored while bis(methylcyclopentadienyl)scandiumchloride ($(MCp)_2ScCl$) was utilized as the Sc precursor. The bubbler temperature utilized in this study ranges between 110-120 °C, which is lower than previously reported (155°C).[18]

To correlate the Sc incorporation with MOCVD growth conditions, a series of samples were designed and compared. Details of growth conditions for three samples (Samples A, B and C) are provided followed by details on four additional samples (Samples D, E, F and G). Samples A and B were grown for 1 hour each while Sample C was grown for 1.5 hrs. For all three samples, the $H_2$ carrier gas flow for the Sc source was set to a constant flow of 500 sccm. Since $(MCp)_2ScCl$ is a relatively new source, there is insufficient data to determine the exact vapor pressure and thus the exact molar flow rate of Sc-precursor estimated to deliver to the reactor. The MO source temperature of the Sc bubbler was set to 110°C for Sample A, while Sample B and C utilized a Sc bubbler temperature of 120°C. Samples A and B utilized a TMAl molar flow rate of 0.53 µmol/min while Sample C utilized a lower TMAl molar flow rate of 0.29 µmol/min.

Three AlScN/AlN/GaN heterostructure samples (Samples D, E and F) with varying AlScN barrier thickness were grown on unintentionally doped (UID) GaN-on-sapphire templates while utilizing the same growth condition as Sample C for the AlScN layer. Each structure comprised of a 300 nm UID GaN layer grown by MOCVD, followed by an AlN interlayer grown for 2 min. The AlScN barrier thickness was varied by adjusting the growth time of the AlScN layer (Sample D: 10 mins, Sample E: 20 mins, Sample F:30 mins). An additional sample (Sample G), using the same AlScN barrier growth condition, was grown on a separate GaN-on-sapphire template directly (without an AlN interlayer) for a duration of 2.5 hrs. This sample was used for scanning transmission electron microscopy (S/TEM) characterization. X-ray diffraction (XRD) 2Θ–ω scans were acquired using a Bruker D8 Discover with a Cu K-α radiation source ($\lambda$ = 1.5418 Å). Surface morphology was characterized by atomic force microscopy (AFM) using a Bruker AXS Dimension system. Mercury probe capacitance–voltage (C–V) measurements were carried out with a Keysight E4980A LCR meter operating at 1 MHz and a mercury vacuum probe contact area of 0.004 cm². Hall measurements were conducted at room temperature on an Ecopia HMS-3000 system equipped with a 0.985 T permanent magnet and indium as ohmic contacts. Cross-sectional scanning transmission electron microscopy (S/TEM) imaging and analysis on selected samples (Sample D and G) were performed to probe the crystalline quality, Sc incorporation and interface properties. Cross-sectional samples for STEM analysis were prepared using a focused ion beam (FIB) lift-out technique in an FEI Helios NanoLab 600 DualBeam system. Specimens were initially thinned with a 5 kV Ga$^+$ ion beam and subsequently polished at 2 kV to minimize surface amorphization and ion-beam damage. High-angle annular dark-field (HAADF) STEM imaging was performed on a Thermo Fisher Scientific Themis Z microscope operated at 200 kV with a semi-convergence angle of 25 mrad. The HAADF detector collection range was 64–200 mrad. To enhance image quality and signal-to-noise ratio, high-resolution HAADF-STEM images were obtained by averaging twenty rapid-scan frames (2048 × 2048 pixels, 200 ns dwell time per pixel). EDS analysis was performed with the Super-X EDS detector, and the elemental mappings are presented as net count images. The Sc compositions within the grown films were also evaluated through bandgap extraction via X-ray photoelectron spectroscopy (XPS). XPS measurements were performed using a ThermoFisher Nexsa G2 system equipped with a scanning XPS microprobe and an Al Kα radiation source (hν = 1486.6 eV). The acquired spectra were calibrated to the C 1s reference peak at 284.5 eV to compensate for surface charging effects. To extract the bandgap of the grown AlScN films, the loss-spectrum of the Al 2p was utilized. The obtained bandgap values were then compared to values shown in literature for sputtered AlScN films to provide estimated Sc compositions of the grown films.[19] Details of the XPS measurements and obtained results are provided in the Supplementary Material.

**Results and Discussions**

2Θ-ω scans of samples A, B and C are shown in Fig. 1. With an increase in the Sc bubbler temperature from 110°C to 120°C, an increase in the Sc incorporation is expected. With the increased Sc concentration for Sample B as compared to Sample A, a shift in the associated film peak is observed. A further shift to the right of the (002) AlScN peak position is observed for Sample C. The TMAl molar flow rate for Sample C was ~45% lower, while the Sc flow rate remained the same as Sample B. For Sample A, its XRD peak appeared at the exact position (36.02°) corresponding to the relaxed (002) AlN. Therefore, it is determined that Sample A has no obvious Sc incorporation due to low Sc flow. XPS analysis of Sample B extracts its Sc composition of ~6% while its 2Θ-ω peak shifts to 36.14°. For Sample C, a further rightward shift in the XRD peak position to 36.18° was observed. The extracted Sc composition of ~13% was obtained from XPS analysis.

The trend in the XRD film peak position shows a rightward shift consistently with an increase in the Sc composition within the range investigated in this work. Shifts in the XRD peak can be affected by the AlScN film composition (and thus resultant lattice constant), relaxation and relative strain state of the film in relation to the underlying substrate, as well as the crystalline quality of the produced film. Literature has shown shifts in the AlScN XRD peak position dependent on growth temperature and thus the thermal strain imparted during growth. Significant rightward shifts were observed with increased growth temperature even with the same Sc composition.[20] Still other work has shown trends of varying peak position shifts with increased Sc composition.[21,22] In this work, we correlate qualitatively of the XRD peak position with the corresponding Sc incorporation in AlScN films. Instead, XPS is used as the primary reference for a quantitative estimation of Sc composition incorporated within the grown films.

Optical microscopy of samples A, B and C reveal a clear trend towards a film with closer lattice-matched conditions for Sample C with higher Sc composition. As shown in Fig. 2, Sample A and Sample B show obvious crack formation which is a result of tensile strain relaxation. The areal density of the formed cracks observed under the same growth duration between Sample A and Sample B suggests a reduced cracking density for Sample B with higher Sc incorporation. In contrast, Sample C does not show obvious crack formation across the surface due to higher Sc composition in this sample. Note that Sample C was grown with longer duration with the targeted film thickness similar to that of Sample A and B. Studies from literature suggest that the lattice matched condition for AlScN is 9%~14% Sc composition rather than the initially theoretically calculated 17%. [12–14]

AFM imaging of 5μm x 5μm scan areas of samples A, B and C are shown in Figs. 3 (a-c). Sample A and B clearly show deep cracks through the film consistent with surface morphology observed from optical microscopy. Sample C shows no obvious cracking across the surface morphology. The root-mean square (Rq RMS) surface roughness of 2.81 nm was measured for a 5μm x 5μm scanned area on Sample C. It should be noted that the surface morphology, while relatively smooth, does not show step-flow growth morphologies typically observed for III-nitrides. This can be related to the relatively narrow growth window in MOCVD of AlScN with the current Sc precursor used which has relatively low vapor pressure and thus low growth rate of the AlScN films.

2Θ-ω scans for AlScN/AlN/GaN heterostructures grown with 3 different AlScN thicknesses are shown in Fig. 4 (Sample D, E, F). From the comparison, as the AlScN thickness increases, the XRD peak shifts toward higher 2Θ angles. The peak intensity of the (002) AlScN increases as the layer grows thicker. In addition, a secondary peak starts to appear at ~35.3° for Sample F which has relatively thick AlScN layer. This could be related to the potential diffusion at the heterointerfaces, e.g., forming AlGaN at the AlN/GaN interface due to extended growth time. AFM surface morphologies of samples D, E and F are shown in Fig. 5. The RMS of surface roughness (Rq RMS) is similar among these samples. There is no obvious surface roughening observable for the samples with thicker AlScN/AlN barrier layers; with all three samples showing Rq RMS ranged from 1.29 nm, up to 1.62 nm.

Fig. 6 shows the mercury probe C-V profiles for all three heterostructure samples (Sample D, E and F). All three samples show sharp transition from the depletion region to the accumulation region in the profiles, indicating high quality GaN/AlN interfaces with 2-dimensional electron gas (2-DEG). As expected, the depletion width increases as the barrier thickness increases from Sample D to Sample F. Assume the area of 0.004 cm$^2$, based on the CV profiles, the extracted 2DEG carrier concentrations are 1.59x10$^{12}$ cm$^{-2}$ (Sample D), 2.84x10$^{12}$ cm$^{-2}$ (Sample E) and 5.22x10$^{12}$ cm$^{-2}$ (Sample F), respectively. However, room temperature Hall measurements of the three samples yielded different values: n ~4.0x10$^{13}$ cm$^{-2}$ with mobility of 780 cm$^2$/Vs (Sample D), and n ~4.8x10$^{13}$ cm$^{-2}$ with mobility of 646 cm$^2$/Vs (Sample E), n ~6.2x10$^{13}$ cm$^{-2}$ with mobility of 562 cm$^2$/Vs (Sample D). The discrepancy between the sheet carrier concentration extracted from mercury C-V probe measurements and Hall measurements for all three samples can be related to several factors: (1) AlScN surface oxidation can lead to a reduced capacitance from CV measurements which in turn underestimates the extracted 2DED carrier concentration; (2) underlying UID GaN layer contributes to the total charge extracted from Hall measurements which overestimates the 2DEG carrier concentration and underestimates the corresponding mobility values; and (3) discrepancy from the estimated contact area sizing from the mercury CV measurements. [23–27] Despite multiple factors that can affect the absolute values of the 2DEG carrier concentrations, the trends from both CV and Hall measurements are consistent. Considering the identical GaN templates were used and the same growth recipe for the regrown GaN buffer layer for the three samples (Sample D, E, and F), the differences in measured sheet carrier concentrations between the three samples from Hall measurements are primarily attributed by the difference in the barrier layer thickness between the samples.

Sample D was selected to perform high-resolution STEM imaging. The cross-sectional imaging at the AlScN/AlN/GaN interfaces is shown in Fig. 7(a). The crystalline structure is well maintained throughout the entire grown layers. The total thickness of the AlScN/AlN barrier layer is ~10 nm, although the interface between the AlScN and AlN is not obvious. Fig. 7(b-i) shows an alternate, larger cross-sectional view field in STEM imaging. STEM-EDS of Al distribution is shown in Fig. 7(b-ii), with a uniform profile of Al throughout the cross-sectional area imaged. A clear interface is also evidenced through the EDS color mapping between the AlN and GaN layers with no apparent Al diffusion into the GaN layer underneath. At the same time, the Sc concentration remains relatively uniform across the top layer of the sample. Considering the allocated growth duration between the AlScN and AlN sublayers (84 % vs. 16 %), it is possible that there exists a transition layer with a graded Sc composition distribution or a delayed incorporation of Sc into the AlScN layer. This phenomenon should be readily detectable for AlScN with higher Sc compositions. STEM imaging was performed for another AlScN film (Sample G), which was grown for an extended duration of 2.5 h with a film thickness of ~70 nm. As shown in Fig. 8(a), an interfacial AlGaN layer was observed, showing different contrast from the GaN and AlScN layers. The extended growth duration likely resulted in alloy formation at the interface. From the STEM-EDS mapping shown in Fig. 8(c), minimal Sc incorporation is observed within this interlayer, suggesting either a delayed Sc incorporation or that the alloyed AlGaN material suppresses Sc diffusion. Interestingly, Sc incorporation, while uniform vertically through the cross-section of the sample, shows some degree of lateral segregation. This behavior can be related to dislocation-type defects that locally affect Sc incorporation. Further investigation is still required to correlate the MOCVD growth conditions and STEM observations.

## Conclusion

In summary, the MOCVD growth of AlScN thin films and AlScN/AlN/GaN heterostructures was systematically investigated using (MCp)$_2$ScCl as the scandium precursor. By controlling the Sc bubbler temperature and adjusting the Al molar flow rate, up to 13% Sc incorporation was achieved with lattice parameters closely matched to GaN. The Al$_{0.87}$Sc$_{0.13}$N/AlN/GaN heterostructures grown on GaN templates exhibited a promising 2DEG channel formed at the AlScN/AlN–GaN interface. The 2DEG carrier density increased with barrier thickness, ranging from $1.59 \times 10^{12}$ to $5.22 \times 10^{12}$ cm$^{-2}$ based on mercury CV measurements, and from $4.0 \times 10^{13}$ to $6.2 \times 10^{13}$ cm$^{-2}$ based on Hall measurements. To accurately determine the 2DEG carrier density and channel mobility, full HEMT device fabrication with the use of a highly resistive GaN buffer layer will be necessary. Overall, the results demonstrate the strong potential of MOCVD for producing high-quality AlScN films and heterostructures suitable for GaN HEMT device applications.

**Supplementary Material**

The supplementary material includes the Al 2p core-level XPS spectra of AlScN films with Sc composition of 6% (Sample B) and 13% (Sample C).

**Acknowledgements**

This work was supported by the Army Research Office (Award No. W911NF-24-2-0210). S.S.-R. acknowledges Dan Huber for assisting with TEM sample preparation. Electron microscopy was performed at the Center for Electron Microscopy and Analysis (CEMAS) at The Ohio State University.

**Conflict of Interest**

The authors have no conflicts of interest to disclose.

**Data Availability**

The data that supports the findings of this study are available from the corresponding author upon reasonable request.

**References**

[1] M. Meneghini, C. De Santi, I. Abid, M. Buffolo, M. Cioni, R.A. Khadar, L. Nela, N. Zagni, A. Chini, F. Medjdoub, G. Meneghesso, G. Verzellesi, E. Zanoni, and E. Matioli, "GaN-based power devices: Physics, reliability, and perspectives," Journal of Applied Physics **130**(18), 181101 (2021).

[2] J.Y. Tsao, S. Chowdhury, M.A. Hollis, D. Jena, N.M. Johnson, K.A. Jones, R.J. Kaplar, S. Rajan, C.G. Van de Walle, E. Bellotti, C.L. Chua, R. Collazo, M.E. Coltrin, J.A. Cooper, K.R. Evans, S. Graham, T.A. Grotjohn, E.R. Heller, M. Higashiwaki, M.S. Islam, P.W. Juodawlkis, M.A. Khan, A.D. Koehler, J.H. Leach, U.K. Mishra, R.J. Nemanich, R.C.N. Pilawa-Podgurski, J.B. Shealy, Z. Sitar, M.J. Tadjer, A.F. Witulski, M. Wraback, and J.A. Simmons, "Ultrawide-Bandgap Semiconductors: Research Opportunities and Challenges," Advanced Electronic Materials **4**(1), 1600501 (2018).

[3] M. Park, Z. Hao, R. Dargis, A. Clark, and A. Ansari, "Epitaxial Aluminum Scandium Nitride Super High Frequency Acoustic Resonators," Journal of Microelectromechanical Systems **29**(4), 490–498 (2020).

[4] Izhar, M.M.A. Fiagbenu, S. Yao, X. Du, P. Musavigharavi, Y. Deng, J. Leathersich, C. Moe, A. Kochhar, E.A. Stach, R. Vetury, and R.H. Olsson, "Periodically poled aluminum scandium nitride bulk acoustic wave resonators and filters for communications in the 6G era," Microsystems & Nanoengineering **11**(1), 19 (2025).

[5] G. Esteves, T.R. Young, Z. Tang, S. Yen, T.M. Bauer, M.D. Henry, and R.H. Olsson III, "Al0.68Sc0.32N Lamb wave resonators with electromechanical coupling coefficients near 10.28%," Applied Physics Letters **118**(17), 171902 (2021).

[6] Y. Zou, C. Gao, J. Zhou, Y. Liu, Q. Xu, Y. Qu, W. Liu, J.B.W. Soon, Y. Cai, and C. Sun, "Aluminum scandium nitride thin-film bulk acoustic resonators for 5G wideband applications," Microsystems & Nanoengineering **8**(1), 124 (2022).

[7] Z. Lu, L. Li, W. Chen, Y. Xiao, W. You, and G. Wu, "A ScAlN-based piezoelectric breathing mode dual-ring resonator with high temperature stability," Microelectronic Engineering **287**, 112144 (2024).

[8] S. Fichtner, N. Wolff, F. Lofink, L. Kienle, and B. Wagner, "AlScN: A III-V semiconductor based ferroelectric," Journal of Applied Physics **125**(11), 114103 (2019).

[9] M.T. Hasan, and Z. Mi, "ScAlN-based HEMTs: Challenges and opportunities," APL Electronic Devices **1**(2), (2025).

[10] H. Wang, S. Mu, and C.G. Van de Walle, "Towards higher electro-optic response in AlScN," Appl. Phys. Lett. **126**(4), (2025).

[11] Z. Li, K. Bian, X. Chen, X. Zhao, Y. Qiu, Y. Dong, Q. Zhong, S. Zheng, and T. Hu, "Demonstration of Integrated AlScN Photonic Devices on 8-Inch Silicon Substrate," Journal of Lightwave Technology **42**(14), 4933–4938 (2024).

[12] D.V. Dinh, J. Lähnemann, L. Geelhaar, and O. Brandt, "Lattice parameters of Sc$_x$Al$_{1−x}$N layers grown on GaN(0001) by plasma-assisted molecular beam epitaxy," Appl. Phys. Lett. **122**(15), 152103 (2023).

[13] L. van Deurzen, T.-S. Nguyen, J. Casamento, H.G. Xing, and D. Jena, "Epitaxial lattice-matched AlScN/GaN distributed Bragg reflectors," Appl. Phys. Lett. **123**(24), (2023).

[14] T.-S. Nguyen, N. Pieczulewski, C. Savant, J.J.P. Cooper, J. Casamento, R.S. Goldman, D.A. Muller, H.G. Xing, and D. Jena, "Lattice-matched multiple channel AlScN/GaN heterostructures," APL Materials **12**(10), 101117 (2024).

[15] S. Leone, J. Ligl, C. Manz, L. Kirste, T. Fuchs, H. Menner, M. Prescher, J. Wiegert, A. Žukauskaitė, R. Quay, and O. Ambacher, "Metal-Organic Chemical Vapor Deposition of Aluminum Scandium Nitride," Physica Status Solidi (RRL) – Rapid Research Letters **14**(1), 1900535 (2020).

[16] S. Krause, I. Streicher, P. Waltereit, L. Kirste, P. Brückner, and S. Leone, "AlScN/GaN HEMTs Grown by Metal-Organic Chemical Vapor Deposition With 8.4 W/mm Output Power and 48 % Power-Added Efficiency at 30 GHz," IEEE Electron Device Letters **44**(1), 17–20 (2023).

[17] C. Manz, S. Leone, L. Kirste, J. Ligl, K. Frei, T. Fuchs, M. Prescher, P. Waltereit, M.A. Verheijen, A. Graff, M. Simon-Najasek, F. Altmann, M. Fiederle, and O. Ambacher, "Improved AlScN/GaN heterostructures grown by metal-organic chemical vapor deposition," Semiconductor Science and Technology **36**(3), 034003 (2021).

[18] I. Streicher, S. Leone, L. Kirste, C. Manz, P. Straňák, M. Prescher, P. Waltereit, M. Mikulla, R. Quay, and O. Ambacher, "Enhanced AlScN/GaN Heterostructures Grown with a Novel Precursor by Metal–Organic Chemical Vapor Deposition," Physica Status Solidi (RRL) – Rapid Research Letters **17**(2), 2200387 (2023).

[19] M. Baeumler, Y. Lu, N. Kurz, L. Kirste, M. Prescher, T. Christoph, J. Wagner, A. Žukauskaitė, and O. Ambacher, "Optical constants and band gap of wurtzite Al$_{1−x}$Sc$_x$N/Al$_2$O$_3$ prepared by magnetron sputter epitaxy for scandium concentrations up to x = 0.41," Journal of Applied Physics **126**(4), 045715 (2019).

[20] P. Wang, D.A. Laleyan, A. Pandey, Y. Sun, and Z. Mi, "Molecular beam epitaxy and characterization of wurtzite Sc$_x$Al$_{1−x}$N," Appl. Phys. Lett. **116**(15), 151903 (2020).

[21] A. Kobayashi, Y. Honda, T. Maeda, T. Okuda, K. Ueno, and H. Fujioka, "Structural characterization of epitaxial ScAlN films grown on GaN by low-temperature sputtering," Appl. Phys. Express **17**(1), 011002 (2023).

[22] D. Solonenko, A. Žukauskaitė, J. Pilz, M. Moridi, and S. Risquez, "Raman Spectroscopy and Spectral Signatures of AlScN/Al2O3," Micromachines **13**(11), 1961 (2022).

[23] E. Tucker, F. Ramos, S. Frey, R.J. Hillard, P. Horváth, G. Zsákai, and A. Márton, "Characterization of AlXGa1−XN/GaN High Electron Mobility Transistor Structures with Mercury Probe Capacitance–Voltage and Current–Voltage," Physica Status Solidi (a) **219**(4), 2100416 (2022).

[24] S.G. Sunkari, H. Das, C. Hoff, Y. Koshka, J.R.B. Casady, and J.B. Casady, "Surface Morphology Improvement and Repeatable Doping Characterization of 4H-SiC Epitaxy Grown on 4° Off-Axis 4H-SiC Wafers," Materials Science Forum **615–617**, 423–426 (2009).

[25] M.J. Uren, D. Lee, B.T. Hughes, P.J.M. Parmiter, J.C. Birbeck, R. Balmer, T. Martin, R.H. Wallis, and S.K. Jones, "Electrical characterisation of AlGaN/GaN heterostructure wafers for high-power HFETs," Journal of Crystal Growth **230**(3), 579–583 (2001).

[26] N. Street, "Materials Development Corporation," (n.d.).

[27] C. Sanna, P. Taylor, R. Hillard, S. Frey, D. McDonald, J. Hoglund, G. Zsakai, A. Marton, and P. Horvath, "Assessment of Surface Preparation Methods for Mercury (Hg) Probe Schottky Capacitance-Voltage (MCV) on Epitaxial Silicon," ECS Journal of Solid State Science and Technology **10**(7), 074006 (2021).

# Figure Captions

**Figure 1:** 2Θ-ω XRD scans for MOCVD grown AlScN films: Sample A (Sc: ~0%), Sample B (Sc: ~6%), and Sample C (Sc: ~13%), grown on GaN-on-sapphire templates. Rightward shift in XRD peak position observable with increased Sc composition.

**Figure 2:** Optical microscopy imaging of AlScN film surfaces of (a) Sample A (Sc: ~0%), (b) Sample B (Sc: ~6%), and (c) Sample C (~13%).

**Figure 3:** AFM imaging of AlScN films of (a) Sample A (Sc: ~0%), (b) Sample B (Sc: ~6%), and (c) Sample C (~13%).

**Figure 4:** 2Θ-ω XRD scans for three AlScN/AlN/GaN heterostructures with varied AlScN/AlN barrier thickness: Sample D (barrier thickness ~10 nm), Sample E (barrier thickness ~20 nm), and Sample F (barrier thickness ~30 nm). Rightward shift in XRD peak position observable with increased barrier thickness. Estimated Sc composition of ~13% for all three samples.

**Figure 5:** AFM imaging of AlScN/AlN/GaN heterostructures for (a) Sample D, (b) Sample E, and (c) Sample F.

**Figure 6:** Mercury probe capacitance-voltage (C-V) curves for the three AlScN/AlN/GaN heterostructures with varying barrier layer thicknesses (Sample D, Sample E and Sample F).

**Figure 7:** (a) Cross-sectional HAADF-STEM imaging of AlScN/AlN/GaN heterostructure (Sample D), (b) (i) larger area cross-sectional STEM imaging, and elemental EDS mapping of the same area for (b) (ii) Al, (b) (iii) Ga, and (b) (iv) Sc.

**Figure 8:** (a) Cross-sectional STEM imaging of Sample G with an AlScN layer grown directly on GaN with extended growth duration (2.5 hrs), (b) HAADF-STEM showing ordered wurtzite crystal structure within the AlScN grown film, and (c) (i) selected region for elemental EDS mapping. Elemental EDS mapping across selected cross-sectional area of (c)(ii) N, (c)(iii) Ga, (c)(iv)Al, (c)(v) Sc, and (c)(vi) Al, Sc combined.

**Figure 1**

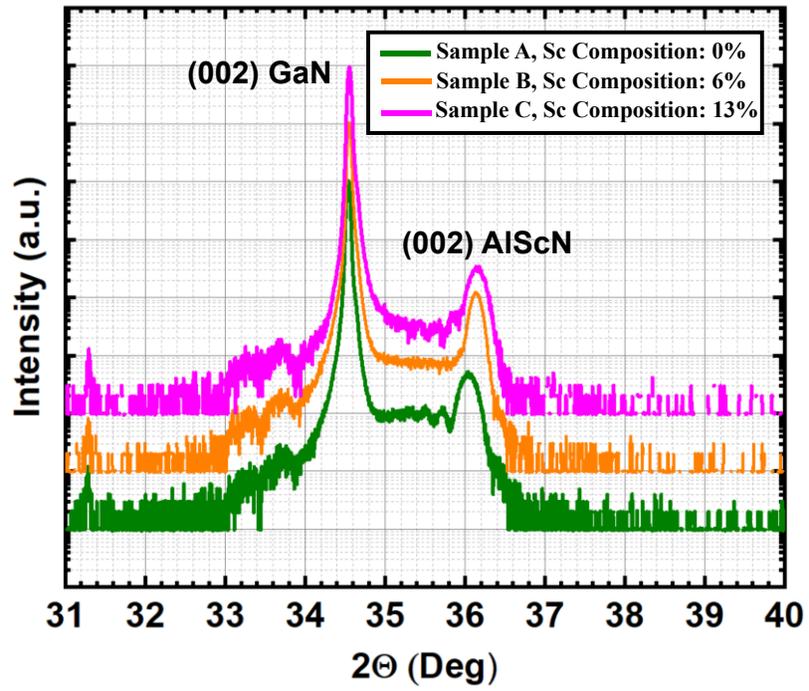

**Figure 2**

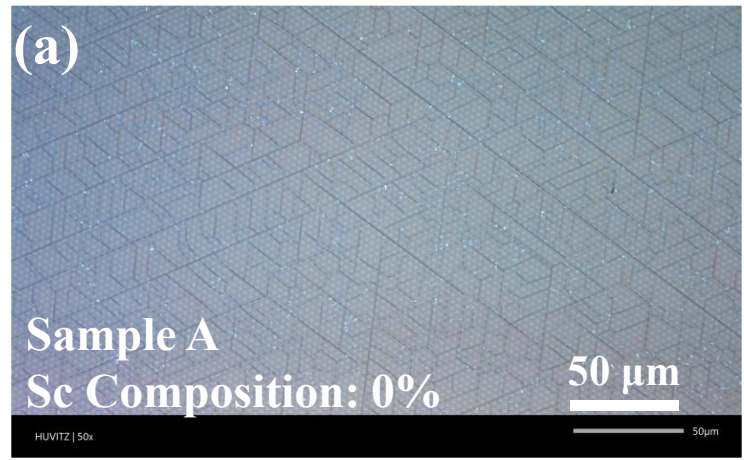

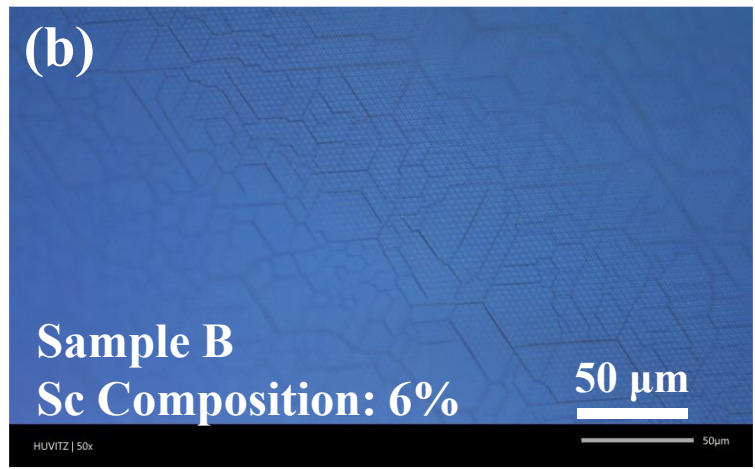

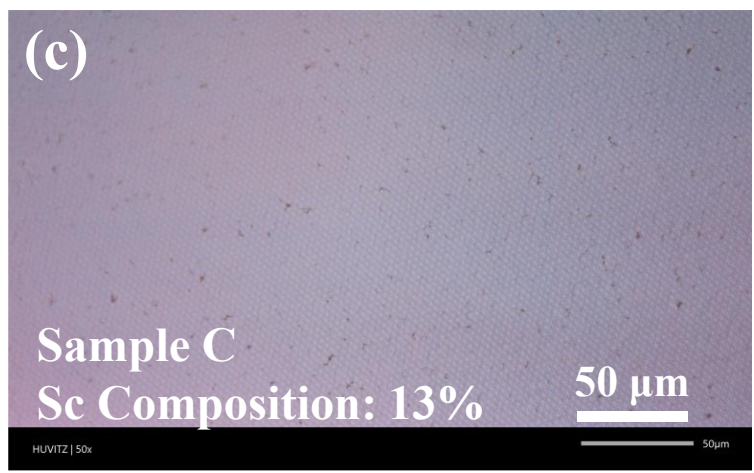

**Figure 3**

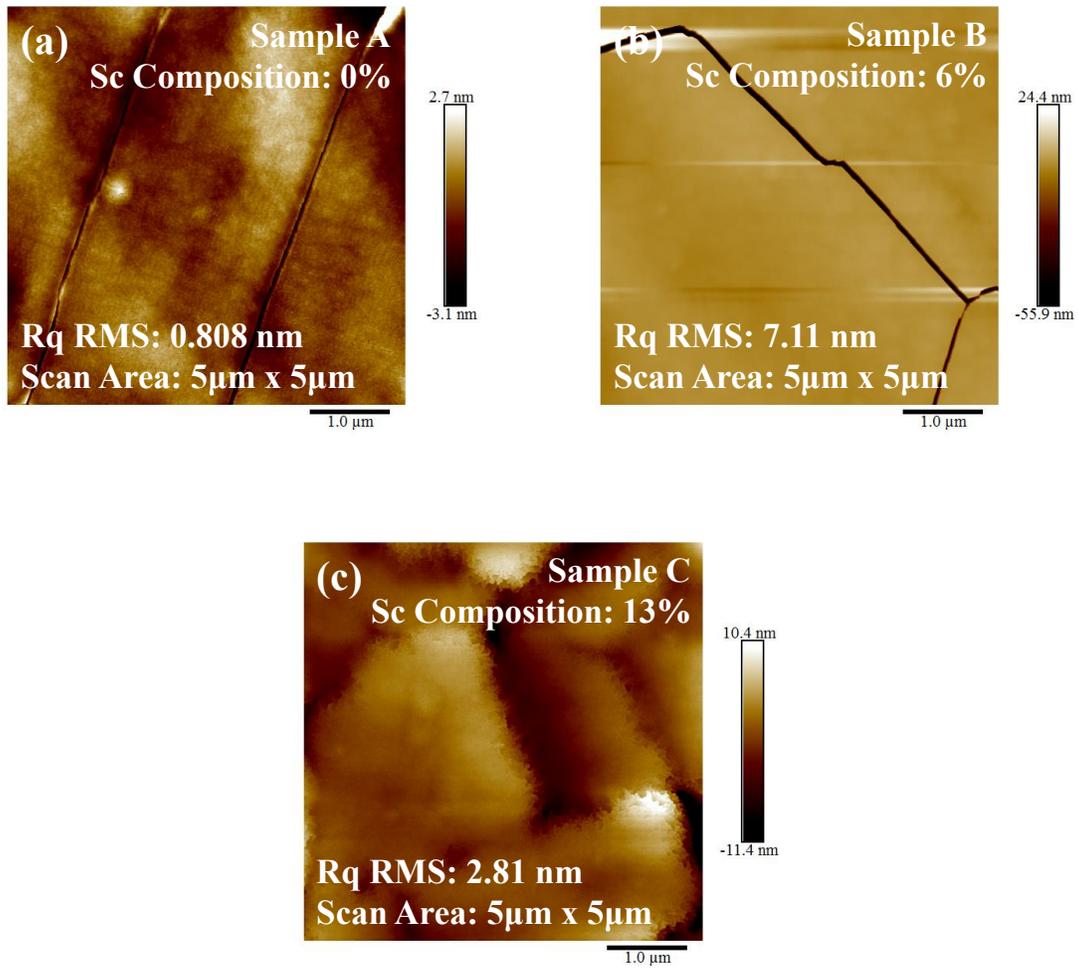

**Figure 4**

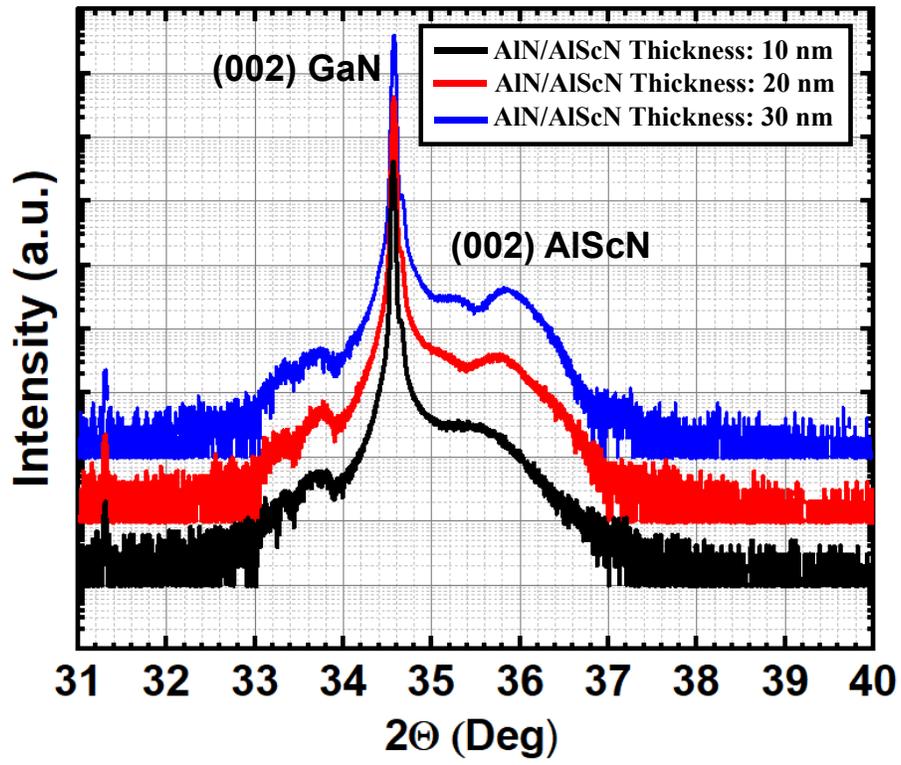

**Figure 5**

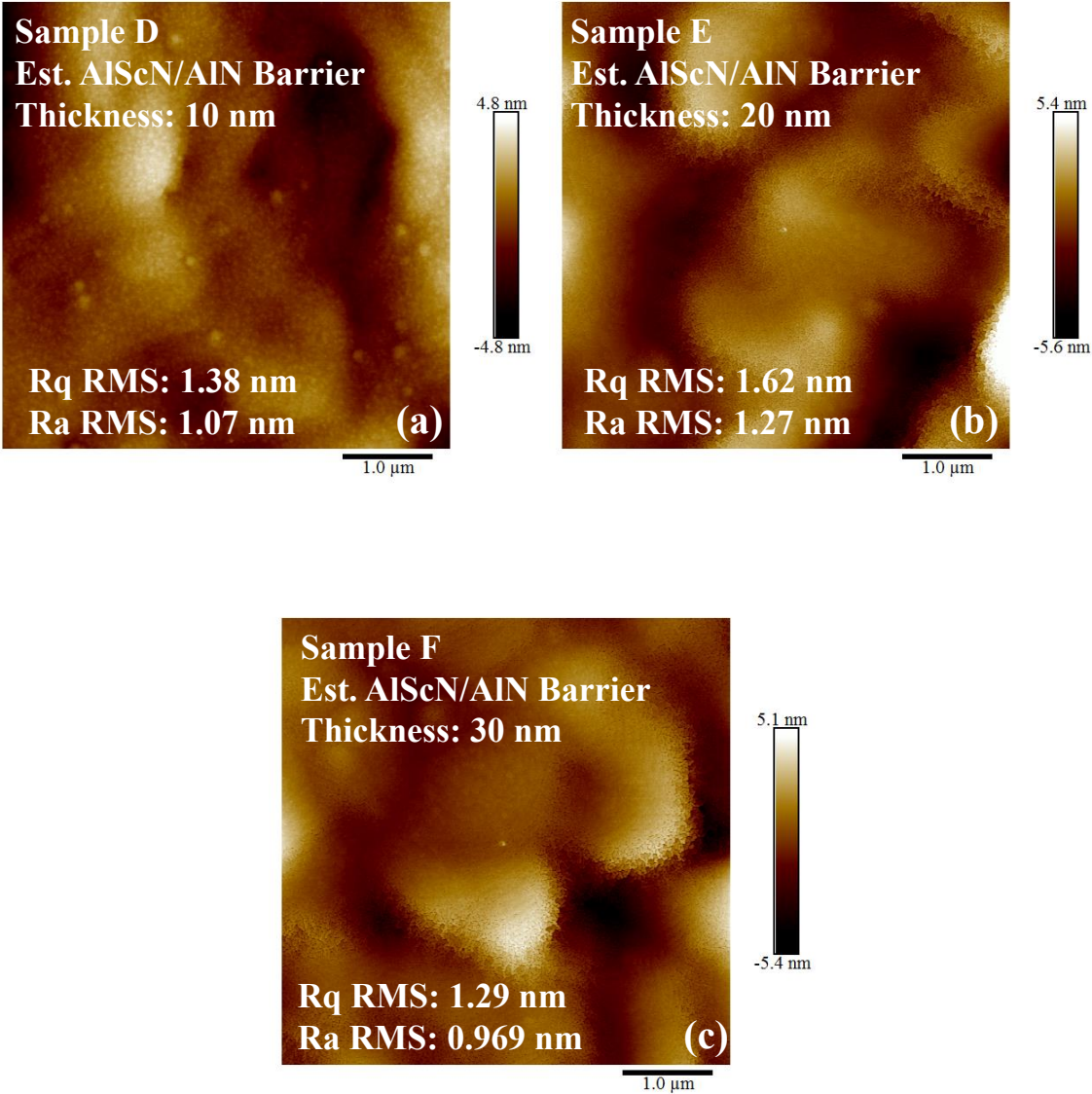

**Figure 6**

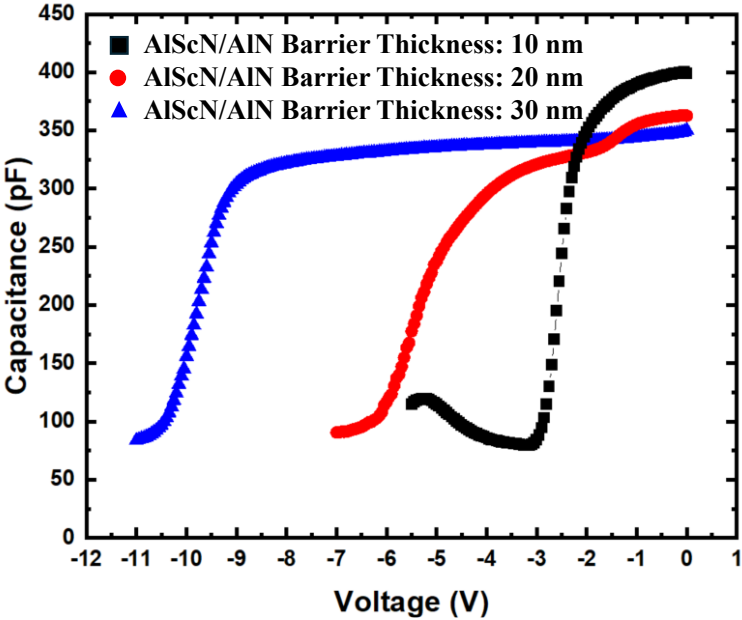

**Figure 7**

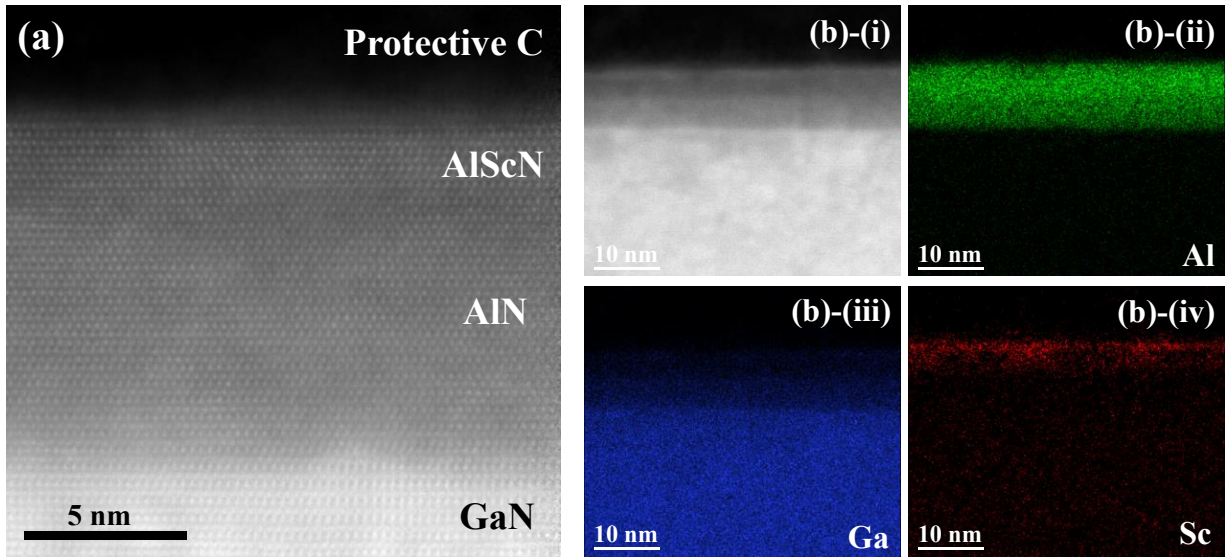

Figure 8

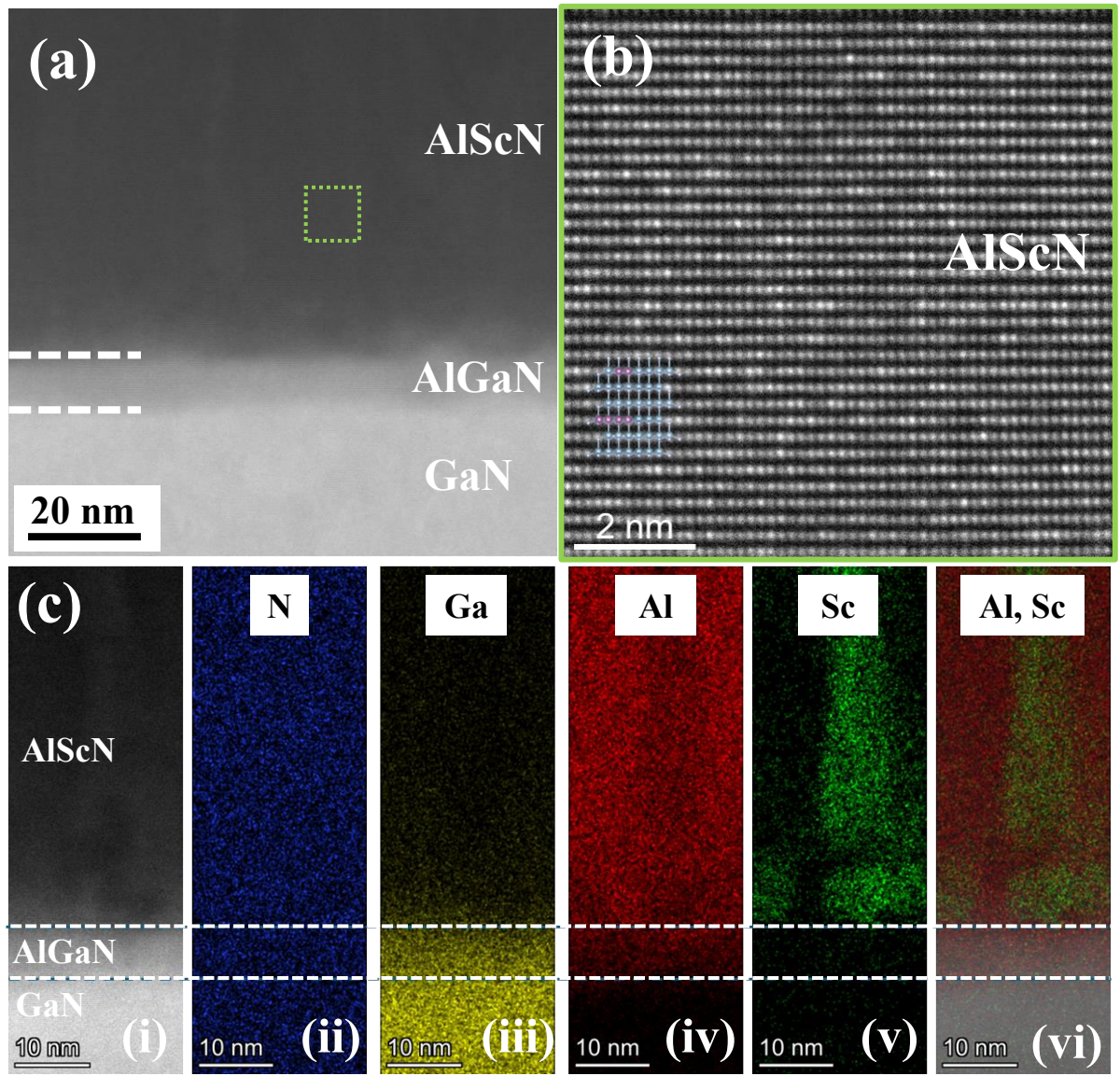

# Supplementary Material

# Metalorganic Chemical Vapor Deposition of AlScN Thin Films and AlScN/AlN/GaN Heterostructures

Vijay Gopal Thirupakuzi Vangipuram[1], Abdul Mukit[1], Kaitian Zhang[1], Salva Salmani-Rezaie[2], Hongping Zhao[1,2,*]

[1] Department of Electrical and Computer Engineering, The Ohio State University, Columbus, Ohio 43210

[2] Department of Materials Science and Engineering, The Ohio State University, Columbus, Ohio 43210

*Corresponding author: zhao.2592@osu.edu

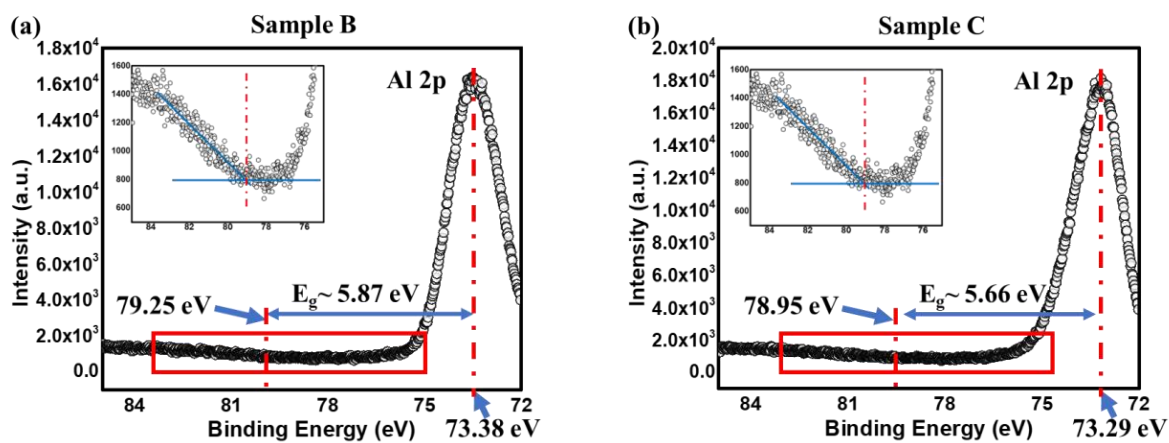

**Figure S1.** Al 2p core-level XPS spectra of $Al_{1-x}Sc_xN$ films with (a) 6 % Sc (Sample B) and (b) 13 % Sc (Sample C), showing the main photoemission peak and the corresponding inelastic energy-loss feature. The onset of the energy-loss tail, indicated by the red dashed lines, was used to estimate the bandgap values.

The bandgap energies of the AlScN films were determined from the inelastic energy-loss features of the Al 2p core-level XPS spectra. As shown in Figures S1 (a) and (b), a distinct inelastic loss tail appears on the higher binding energy side of the Al 2p peak for both samples. The onset of this loss feature, obtained by linear fitting of the rising edge (indicated by the red dash-doted lines), corresponds to the bandgap energy of the material. For the film with 6 % Sc (Sample B), the extracted bandgap is **5.87 ± 0.04 eV**, whereas the film with 13 % Sc (Sample C) exhibits a slightly narrower bandgap of **5.66 ± 0.04 eV**. This clear reduction in bandgap with increasing Sc incorporation is consistent with the theoretical predictions for AlScN alloys. The rate of decrease (≈ 39 meV per at.% Sc) aligns closely with the reported theoretical and experimental trends for the low-Sc regime, confirming that moderate Sc incorporation effectively tunes the electronic structure while retaining the ultrawide-bandgap nature of the material.